\documentclass[aps,prl,amsfonts,preprintnumbers,superscriptaddress,showpacs,nofootinbib,twocolumn]{revtex4}

\usepackage{graphics}
\usepackage{psfrag}
\usepackage{epsfig}
\usepackage{psfig}
\usepackage{epsf}
\usepackage{float}

\newcommand{\fet}[1]{\mbox{\boldmath $#1$}}

\newcommand{\beq}{\begin{equation}}
\newcommand{\eeq}{\end{equation}}
\newcommand{\beqa}{\begin{eqnarray}}
\newcommand{\eeqa}{\end{eqnarray}}
\newcommand{\nn}{\nonumber \\ }

\newcommand{\3}{{\ss}}

\begin{document}

\preprint{JLAB-THY-05-440}

\title{Four--nucleon force in chiral effective field theory}

\author{E. Epelbaum}
\email[]{Email: epelbaum@jlab.org}
\affiliation{Jefferson Laboratory, Theory Division, Newport News, VA 23606, USA}

\date{\today}

\begin{abstract}
We derive the leading contribution to the four--nucleon force within the framework of 
chiral effective field theory. It is governed by the exchange of pions and the lowest--order 
nucleon--nucleon contact interaction and includes effects due to the nonlinear  
pion--nucleon couplings and the pion self interactions constrained by the chiral 
symmetry of QCD. The resulting four--nucleon force does not contain any unknown parameters and can be tested
in few-- and many--nucleon studies. 
\end{abstract}


\maketitle

\vspace{-0.2cm}
Chiral effective field theory (EFT) is a powerful tool for analyzing the properties of 
hadronic systems at low energies in a systematic and model--independent way. 
It exploits the approximate and spontaneously broken chiral symmetry of QCD 
which governs low--energy hadron structure and dynamics. Over the past years, considerable progress 
has been achieved in understanding the structure of the nuclear force in 
this framework \cite{Weinberg:1990rz,Weinberg:1991um}, see \cite{Epelbaum:review} for a review.
In particular, two-- (2NFs) 
and three--nucleon forces (3NFs) have been worked out and applied in few--nucleon studies 
upto next--to--next--to--next--to--leading (N$^3$LO) \cite{Entem:2003ft,Epelbaum:2004fk} 
and next--to--next--to--leading orders (N$^2$LO) \cite{Epelbaum:2002vt} in the chiral expansion, respectively.  
Parallel to these developments, significant progress has been achieved in the microscopic description of 
few-- and many--nucleon systems. Advances in the development of methods, such as 
the Faddeev--Yakubovsky scheme, Green's function Monte Carlo and hyperspherical harmonics methods
as well as the no--core shell model, coupled with a significant increase in computational resources 
allow one to obtain the 
binding energies of light nuclei with high numerical accuracy reaching 0.2\% 
for the $\alpha$--particle \cite{Nogga:2001cz} and 1-2\% for heavier nuclei
\cite{Navratil:2003ef,Pieper:2002ne}. Three--  
and four--nucleon  continuum observables are also being investigated.  
These studies open the door for precise and nontrivial tests of the underlying dynamics. 
In this context, three--nucleon forces  have deserved much attention and 
were shown to be important e.g.~for understanding the spectra of light nuclei \cite{Pieper:2002ne}. 
Four--nucleon forces (4NFs) have not yet been explored in few--body calculations. 
Although the existing studies leave little room for the contribution of the possible 4NFs,
see e.g.~\cite{Nogga:2001cz}, this needs to be verified via explicit calculations. 
In this paper we make a step in this direction and derive the leading 4NF in the 
framework of chiral EFT which appears at N$^3$LO in the chiral expansion. The resulting 4NF is local and parameter--free. 
Its application in future few--body studies will provide a challenging test of our
understanding of the nature of nuclear forces. 

Let us first recall the structure 
of the effective chiral Lagrangian for pions and nucleons, which 
at lowest order reads:
\beqa
\label{lagr}
\mathcal{L}^{(2)}_{\pi \pi} &=& \frac{F_\pi^2}{4} \, \mbox{tr}  \left[ \partial_\mu U \partial^\mu U^\dagger
+ M_\pi^2 ( U + U^\dagger ) \right]\,,\\
\mathcal{L}^{(1)}_{\pi N} &=& N^\dagger \left( i D_0 - \frac{g_A}{2} \vec \sigma \cdot \vec u \right) N \,,\nn
\mathcal{L}^{(0)}_{NN} &=& -\frac{1}{2} C_S ( N^\dagger N )  ( N^\dagger N )  - \frac{1}{2} C_T ( N^\dagger \vec \sigma N ) 
\cdot ( N^\dagger \vec \sigma N )\,. 
\nonumber
\eeqa
Here the superscripts refer to the number of derivatives and/or $M_\pi$--insertions, $F_\pi = 92.4$ MeV is the 
pion decay constant and $g_A = 1.267$ is the nucleon axial--vector coupling. The low--energy constants (LECs)
$C_S$ and $C_T$ determine the strength of the leading NN short--range interaction  
\cite{Weinberg:1990rz,Weinberg:1991um}. The Lagrangians $\mathcal{L}^{(1)}_{\pi N}$ and $\mathcal{L}^{(0)}_{NN}$
are given in the heavy--baryon formulation with $N$ representing a non--relativistic nucleon field and 
$\vec \sigma$ denoting the Pauli spin matrices. Further, the SU(2) matrix $U = u^2$ collects the pion fields,
$D^\mu = \partial^\mu + \frac{1}{2} [ u^\dagger , \, \partial^\mu u ]$ denotes the covariant derivative of 
the nucleon field and $u_\mu = i u^\dagger \partial_\mu U u^\dagger$. 
The first terms in the expansion of the matrix $U(\fet \pi )$ in powers of the pion fields 
take the form, see e.g.~\cite{Kaiser:1999ff}:
\beq
U (\fet \pi ) = 1 + \frac{i}{F_\pi} \fet \tau \cdot \fet \pi - \frac{1}{2 F_\pi^2} \fet \pi^2 - \frac{i \alpha}{F_\pi^3} 
(\fet \tau \cdot \fet \pi )^3 + \frac{8 \alpha - 1}{8 F_\pi^4} \fet \pi^4 + \ldots\,,
\eeq
where $\fet \tau$ denote the Pauli isospin matrices and $\alpha$ is an arbitrary constant. 
Notice that only the coefficients in front of the linear and quadratic terms in the pion field are 
fixed uniquely from the unitarity condition $U^\dagger U=1$ and the proper normalization of the 
pion kinetic energy. The explicit $\alpha$--dependence of the matrix $U$
represents the freedom in the definition of the pion field. Clearly, all measurable quantities are
$\alpha$--independent. 

The structure of the nuclear forces can be visualized utilizing a diagrammatic language. 
Given the non--uniqueness of the nuclear potential, the precise meaning of diagrams 
is scheme--dependent. Typical examples include Feynman graphs with subtracted iterative 
contributions, see e.g.~\cite{Kaiser:1999ff}, irreducible time--ordered diagrams \cite{Weinberg:1990rz,Weinberg:1991um} 
or time--ordered--like graphs in the method of unitary transformation \cite{Epelbaum:review}. 
The importance of a particular contribution to the nuclear force is determined 
by the power $\nu$ of the expansion 
parameter $Q/\Lambda$. Here, $Q$ and $\Lambda$ refer to the generic low--momentum scale associated with 
external nucleon momenta or $M_\pi$ and the pertinent hard scale, respectively.
For a $N$--nucleon diagram with  $L$ loops, $C$ separately connected pieces
and $V_i$ vertices of type $i$, the low--momentum dimension is given by \cite{Weinberg:1990rz,Weinberg:1991um}
\beqa
\label{powc}
\nu &=& -2 + 2 N - 2 C + 2 L + \sum_i V_i \Delta_i \\
&& \mbox{with} \quad \Delta_i = d_i + \frac{1}{2} n_i - 2\,.
\nonumber
\eeqa
Here, $n_i$ is the number of nucleon field operators and $d_i$ the  
number of derivatives and/or insertions of $M_\pi$.
The interaction terms of lowest possible chiral dimension $\Delta_i =0$ are listed 
in Eq.~(\ref{lagr}).
The power counting in Eq.~(\ref{powc}) naturally explains the dominance of the two--nucleon force (2NF)
which starts to contribute at order $\nu = 0$ \cite{Weinberg:1990rz,Weinberg:1991um}. The leading 3NF  
formally arises at order $\nu = 2$. The corresponding contributions are known to  
vanish provided one uses an energy--independent formulation such as the method of unitary 
transformation \cite{Coon:1986kq,Eden:1996ey,Epelbaum:1998ka}, see also 
\cite{Yang:1986pk,vanKolck:1994yi} for a similar conclusion based on the framework of time--ordered 
perturbation theory. The first non-vanishing 3NFs appear at order  $\nu = 3$ 
\cite{vanKolck:1994yi,Epelbaum:2002vt}. At present, the 2NFs and 3NFs have been worked out and applied 
in few--body studies upto orders $\nu = 4$ \cite{Entem:2003ft,Epelbaum:2004fk} and $\nu = 3$
\cite{Epelbaum:2002vt}, respectively, see \cite{Epelbaum:review} for more details.
Derivation of the leading 3NF corrections at order   $\nu = 4$ is in progress.

\begin{figure}[tb]
\includegraphics[width=4.3cm,keepaspectratio,angle=0,clip]{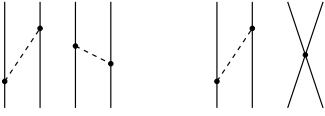}
    \caption{
         Disconnected diagrams at order $\nu = 2$ that lead to vanishing contributions to 
         the four--nucleon force. Solid and dashed lines represent nucleons and pions,
         respectively. 
\label{fig1} 
 }
\end{figure}

\begin{figure}[tb]
\includegraphics[width=8.5cm,keepaspectratio,angle=0,clip]{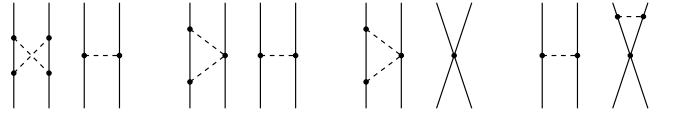}
    \caption{Examples of disconnected diagrams at order $\nu = 4$ that lead to vanishing contributions to 
         the four--nucleon force. 
\label{fig4} 
 }
\end{figure}

\begin{figure}[tb]
\includegraphics[width=8.5cm,keepaspectratio,angle=0,clip]{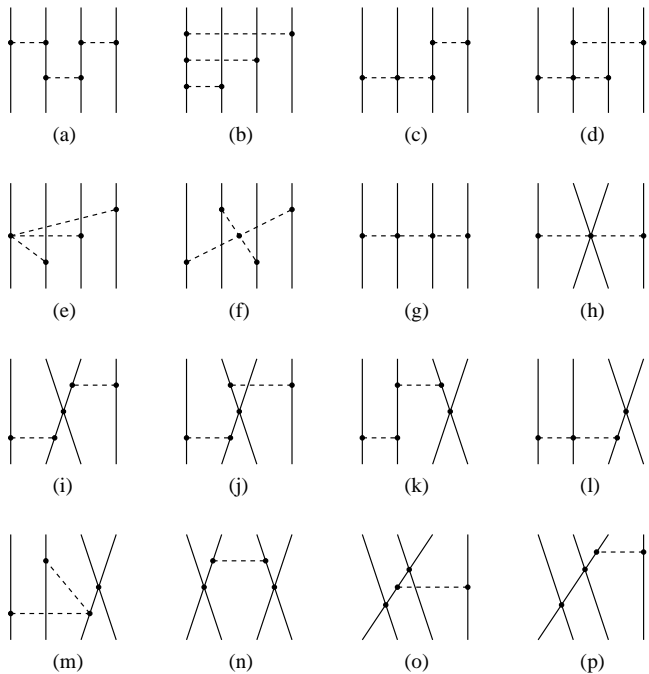}
    \caption{
         The leading contributions to the four--nucleon force. 
         Graphs resulting by the interchange of the vertex ordering and/or nucleon lines are not shown.
\label{fig2} 
 }
\end{figure}

According to Eq.~(\ref{powc}), the leading 4NF is expected to arise at order $\nu = 2$ from  disconnected tree
diagrams with the lowest--order vertices, see Fig.~\ref{fig1}. 
Using the method of unitary transformation, 
it is easy to see that both graphs in Fig.~\ref{fig1} yield vanishing contributions,
see also \cite{vanKolck:1994yi} for an earlier discussion based on time--ordered perturbation theory. 
The first non-vanishing 4NFs appear at order $\nu = 4$ where  
one has to take into account disconnected tree diagrams with one insertion of a $\Delta_i = 2$ vertex as well as 
disconnected loop and connected tree graphs with the lowest--order vertices depicted in Figs.~\ref{fig4} and \ref{fig2}, respectively. 
In addition, one needs to consider  
relativistic $1/m$--corrections to disconnected diagrams shown in Fig.~\ref{fig1}. We have verified that {\it all} disconnected 
graphs at this order yield vanishing contributions to the 4NF. One is, therefore, left with connected tree diagrams 
shown in Fig.~\ref{fig2}.  
We have used the method of unitary transformation \cite{Epelbaum:1998ka} to evaluate the corresponding 4NF contributions. 
To that aim we first switch from the 
Lagrangian in Eq.~(\ref{lagr}) to the effective $\pi N$ Hamilton operator using the canonical formalism and then apply
a suitable unitary transformation in Fock space 
in order to decouple purely nucleonic states from the ones which contain pions.
The resulting Hamilton operator acts on the nucleon subspace of the Fock space and  
provides a definition of the nuclear forces. A rather tedious 
calculation along the lines of Ref.~\cite{Epelbaum:1998ka} yields the following result
for the contributions of individual diagrams in Fig.~\ref{fig2}:
\beqa
\label{4nf}
V^{a} &=& - \frac{2 g_A^6}{( 2 F_\pi )^6} 
\frac{\vec \sigma_1 \cdot \vec q_1 \;\vec \sigma_4 \cdot \vec q_4}{[\vec q_1^{\;2}  + M_\pi^2]\,
[\vec q_{12}^{\;2}  + M_\pi^2]^2 \, [\vec q_4^{\;2}  + M_\pi^2]} \nn
&\times& \Big[ ( \fet \tau_1 \cdot \fet \tau_4 \,  \fet \tau_2 \cdot \fet \tau_3
-  \fet \tau_1 \cdot \fet \tau_3 \,  \fet \tau_2 \cdot \fet \tau_4 ) \,\vec q_1 \cdot \vec q_{12}
\, \vec q_4 \cdot \vec q_{12} \nn
&& {} + \fet \tau_1 \times \fet \tau_2 \cdot \fet \tau_4  \; \vec q_1 \cdot \vec q_{12}  \; 
\vec q_{12} \times \vec q_4 \cdot \vec \sigma_3 \nn
&& {}  + \fet \tau_1 \times \fet \tau_3 \cdot \fet \tau_4  \; \vec q_4 \cdot \vec q_{12}  \; 
\vec q_{1} \times \vec q_{12} \cdot \vec \sigma_2  \nn
&& {} + \fet \tau_1 \cdot \fet \tau_4 \; \vec q_{12} \times \vec q_{1} \cdot \vec \sigma_2 \;
\vec q_{12} \times \vec q_4 \cdot \vec \sigma_3  \Big] +\mbox{all perm.},\nn [4pt]
V^{c} &=& - \frac{2 g_A^4}{(2 F_\pi)^6} 
\frac{\vec \sigma_1 \cdot \vec q_1 \;\vec \sigma_4 \cdot \vec q_4}{[\vec q_1^{\; 2}  + M_\pi^2]\,
[\vec q_{12}^{\; 2}  + M_\pi^2]\, [\vec q_4^{\; 2}  + M_\pi^2]}  \nn
&& {} \times \Big[ (  \fet \tau_1 \cdot \fet \tau_4 \,  \fet \tau_2 \cdot \fet \tau_3
-  \fet \tau_1 \cdot \fet \tau_3 \,  \fet \tau_2 \cdot \fet \tau_4 ) \, \vec q_{12} \cdot \vec q_4 \nn
&& {} +  \fet \tau_1 \times \fet \tau_2 \cdot \fet \tau_4  \; \vec q_{12} \times \vec q_4 \cdot \vec \sigma_3 \Big] +\mbox{all perm.}, \nn [4pt]
V^{e} &=& \frac{g_A^4}{(2 F_\pi)^6} 
\frac{\vec \sigma_2 \cdot \vec q_2 \;\vec \sigma_3 \cdot \vec q_3 \;\vec \sigma_4 \cdot \vec q_4}{[\vec q_2^{\;2}  + M_\pi^2]\,
[\vec q_{3}^{\;2}  + M_\pi^2] \, [\vec q_4^{\;2}  + M_\pi^2]}  \nn [1pt]
&& {} \times \fet \tau_1 \cdot \fet \tau_2 \,  \fet \tau_3 \cdot \fet \tau_4 \; 
\vec \sigma_1 \cdot (\vec q_3 + \vec q_4 ) + \mbox{all perm.}, \nn   [4pt]
V^{f} &=& \frac{g_A^4}{2 (2 F_\pi)^6} 
\; \Big[ \left( \vec q_1 + \vec q_2 \, \right)^2 + M_\pi^2 \Big] \nn
&& {} \times 
\frac{\vec \sigma_1 \cdot \vec q_1 \;\vec \sigma_2 \cdot \vec q_2 \;\vec \sigma_3 \cdot \vec q_3 \;\vec \sigma_4 \cdot \vec q_4}
{[\vec q_1^{\;2}  + M_\pi^2]\,  [\vec q_{2}^{\;2}  + M_\pi^2] \, [\vec q_{3}^{\;2}  + M_\pi^2] \, [\vec q_4^{\;2}  + M_\pi^2]} \nn  [2pt]
&& {} \times  \fet \tau_1 \cdot \fet \tau_2 \,  \fet \tau_3 \cdot \fet \tau_4   + \mbox{all perm.,} \nn [4pt]
V^{k} &=& 4 C_T \frac{g_A^4}{(2 F_\pi)^4} \,
\frac{\vec \sigma_1 \cdot \vec q_1 \; \vec \sigma_3 \times \vec \sigma_4  \cdot \vec q_{12}}
{[\vec q_1^{\;2}  + M_\pi^2]\, [\vec q_{12}^{\;2}  + M_\pi^2]^2} \nn
&& {} \times  \Big[ \fet \tau_1 \cdot \fet \tau_3 \; \vec q_1 \times \vec q_{12} \cdot \vec \sigma_2   - 
\fet \tau_1 \times \fet \tau_2 \cdot \fet \tau_3 \; \vec q_{1} \cdot \vec q_{12} \Big] \nn [1pt]
&& {} +  \mbox{all perm.}, \nn [4pt]
V^l &=& - 2 C_T \frac{g_A^2}{(2 F_\pi)^4} \,
\frac{\vec \sigma_1 \cdot \vec q_1 \; \vec \sigma_3 \times \vec \sigma_4  \cdot \vec q_{12}}
{[\vec q_1^{\;2}  + M_\pi^2] \, [\vec q_{12}^{\;2}  + M_\pi^2]} \; \fet \tau_1 \times \fet \tau_2 \cdot \fet \tau_3   \nn [2pt]
&& {} + \mbox{all perm.}, \nn[4pt]
V^n &=&  2 C_T^2 \frac{g_A^2}{(2 F_\pi)^2} \,
\frac{\vec \sigma_1 \times \vec \sigma_2  \cdot \vec q_{12} \; \vec \sigma_3 \times \vec \sigma_4  \cdot \vec q_{12}}
{[\vec q_{12}^{\;2}  + M_\pi^2]^2} \;  \fet \tau_2 \cdot \fet \tau_3  \nn
&& {} + \mbox{all perm.} . 
\eeqa 
Here, the subscripts refer to the nucleon labels and $\vec q_{i} = \vec p_i \, ' - \vec p_i$ with $\vec p_i \, '$
and $\vec p_i$ being the final and initial momenta of the nucleon i. 
Further, $\vec q_{12} = \vec q_1 + \vec q_2 = - \vec q_3 - \vec q_4 = -\vec q_{34}$ is the momentum transfer between the 
nucleon pairs 12 and 34. Notice that some terms in the sum over all possible 24 permutations of the nucleon labels 
are the same. Diagrams (b), (d), (g), (h), (i), (j), (m), (o) and (p) lead to vanishing contributions  
to the 4NF. Notice further that the short--range 4NFs only depend on the LEC $C_T$ which can be traced back to the fact that 
the $C_S$--vertex commutes with the other lowest--order vertices.  Details of the derivation 
will be published elsewhere.

The results given in Eq.~(\ref{4nf}) deserve some special comments. 
The easiest way to evaluate the contributions from graphs (e) and (f), which do not include reducible topologies 
(i.~e.~time--ordered diagrams with purely nucleonic intermediate states), is by calculating the corresponding 
Feynman diagrams. The resulting 4NFs depend individually on the parameter $\alpha$ which enters the effective Lagrangian   
in Eq.~(\ref{lagr}). When they are added together, this $\alpha$--dependence cancels out, see also 
\cite{Mcmanus:1980ep,Robilotta:1985gv} for an earlier discussion of these contributions. 
We only show explicitly  the remaining, $\alpha$--independent terms 
in Eq.~(\ref{4nf}).
The $3\pi$--exchange 4NF proportional to $g_A^2$ can be obtained by evaluating the Feynman diagram 
(g) in Fig.~\ref{fig2}. Due to the four--momentum conservation at each vertex and the fact that 
the Weinberg--Tomozawa vertex contains a time derivative of the pion field, this 
contribution is suppressed by $Q^2/m^2$ and, therefore, shifted to higher 
orders. In the method of unitary transformation, one needs to consider not only graph (g) but also 
graph (h), where the $\pi \pi NNNN$ vertex of dimension 
$\Delta_i =0$,
\beq
\mathcal{H}_{\rm int} = \frac{1}{2 (2 F_\pi )^4} \,( N^\dagger \fet \tau \times \fet \pi  N ) 
\cdot ( N^\dagger \fet \tau \times \fet \pi  N )\,,
\eeq
arises in the Hamiltonian through the application of the canonical formalism to the 
Lagrangian in Eq.~(\ref{lagr}), see e.g.~\cite{Weinberg:1992yk}. We have verified
that the 4NF contributions from diagrams (g) and (h) in this framework cancel against each other, 
which is consistent with the result based on the Feynman graph technique.  

\begin{figure}[tb]
\includegraphics[width=6.7cm,keepaspectratio,angle=0,clip]{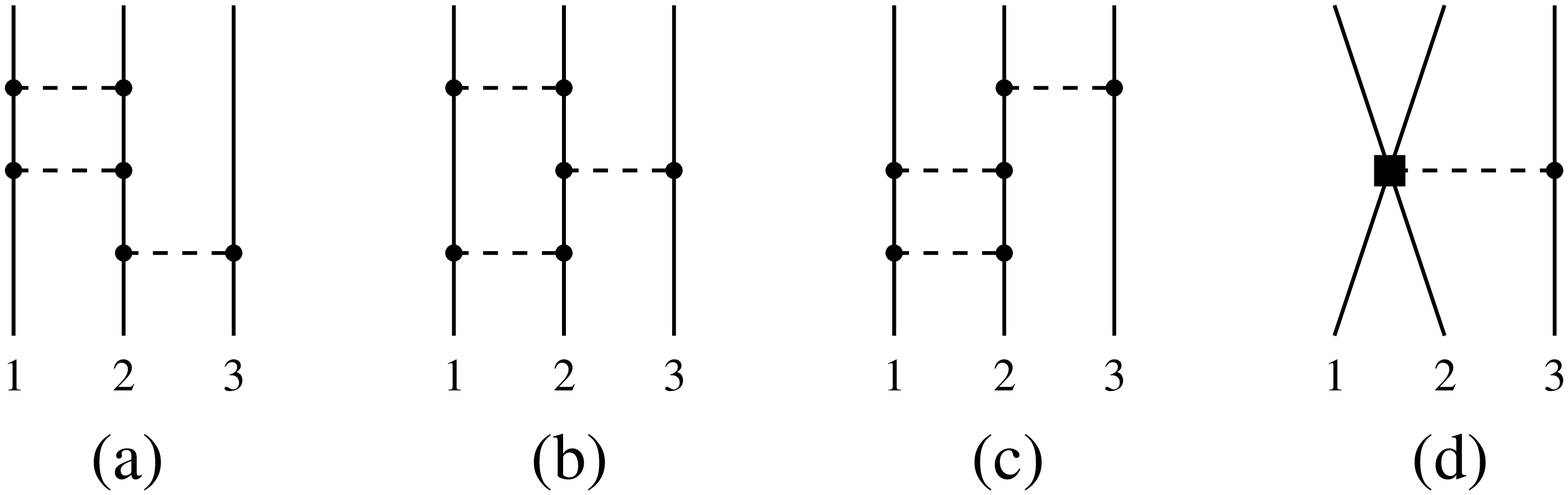}
    \caption{
         Selected contributions to the three--nucleon force.
\label{fig3} 
 }
\end{figure}

Before discussing the numerical size of the derived 4NF, another important comment is in order.
Nuclear forces are not unique and can, in general, be changed via unitary transformations. We 
have explored a large class of unitary transformations which act on the nucleon subspace 
of the Fock space and affect nuclear forces at order $\nu = 4$ (and higher). 
We found that the associated ambiguity in the definition of the nuclear forces is resolved 
if one requires renormalizability of the 1--loop 3NF contributions.
Consider, for example, $3\pi$--exchange 3NF corresponding to  
diagrams (a--c) in Fig.~\ref{fig3}. 
Ultraviolet--divergent  loop integrals resulting from $2 \pi$--exchange between the nucleons 1 and 2
have to be renormalized by a redefinition of the 
LEC $D$ which enters the expression for the 3NF at order $\nu = 3$ \cite{vanKolck:1994yi,Epelbaum:2002vt}, see diagram (d) 
in Fig.~\ref{fig3}:
\beq
\label{3nf}
V_{3N} = - D \frac{g_A}{8 F_\pi^2}   \frac{\vec \sigma_2 \cdot \vec q_3 \; \vec \sigma_3 \cdot \vec q_3}
{[ \vec q_3^{\; 2} + M_\pi^2]} \fet \tau_2 \cdot \fet \tau_3 +   \mbox{all perm.}.
\eeq
For this to be possible, ultraviolet--divergent pieces in the $3\pi$--exchange 3NF must have the same 
form as $V_{3N}$. As a consequence, no terms proportional 
to $( \vec q_3^{\; 2} + M_\pi^2)^{-2}$, $( \vec q_3^{\; 2} + M_\pi^2)^{-3/2}$ or
$(\sqrt{\vec q_3^{\; 2} + M_\pi^2} + \sqrt{\vec l_1^{\; 2} + M_\pi^2} )^{-1}$ with $\vec l_1$
being the momentum of one of the two pions exchanged between the nucleons 1 and 2 should appear in the 3NF. 
This restriction imposes nontrivial constraints on the choice of the unitary transformations and, therefore, also 
on the form of the 4NFs corresponding to graphs (a) and (b) in Fig.~\ref{fig2}.
These considerations may remind one of the recent 
findings in the context of large $N_c$ QCD \cite{Cohen:2002im}. 
One should, of course, keep in mind that S--matrix elements, which involve iterations of the 
potential, are not affected by unitary transformations. 
Based on the above arguments, we have not found any ambiguity in the form of the 4NFs 
(for the considered class of unitary transformations). We postpone a detailed 
discussion of this issue to a forthcoming publication. 

In order to test the effects of the 4NFs in few--nucleon systems, explicit calculations will need  
to be performed. To get a rough idea about the size of the 4NF contributions to e.g.~the $\alpha$--particle binding energy,
one can look at the strength of the corresponding $r$--space potentials. For $V^e$, for example, it is given by
\beq
\frac{1}{(4 \pi)^3} \,  \frac{2 g_A^4}{( 2 F_\pi)^6} M_\pi^7 \sim 80 \mbox{ keV}\,,
\eeq  
see also \cite{Robilotta:1985gv}. 
Here, the factor $(4 \pi)^{-3}$ arises from integrations over the relative momenta. Other $3\pi$--exchange 4NFs have similar strengths. 
The short--range terms can be expected to be less important due to numerically small
values of the LEC $C_T$ found in \cite{Epelbaum:2001fm} and (for certain cut--off combinations) in 
\cite{Epelbaum:2004fk}. We remind the reader that in EFT without or with perturbative pions, 
one has $C_T=0$ in the limit when both NN S--wave scattering lengths go to infinity \cite{Mehen:1999qs}.

To summarize, we have derived the leading 4NF contributions based on chiral EFT that appear 
at N$^3$LO in the low--momentum expansion. 
Chiral symmetry of QCD  fixes the structure and strength of the 
pion couplings and plays a crucial role in the derivation. 
The strength of the lowest--order NN contact interaction entering the expressions for the 4NF
is determined in the two--nucleon system \cite{Entem:2003ft,Epelbaum:2004fk}, so that the resulting 4NF is 
parameter--free. Applications of this 4NF to study the properties of the $\alpha$--particle and the spectra of other 
light nuclei will provide a challenging test of our understanding of the nuclear force.

The author would like to thank Sid Coon, Walter Gl\"ockle, Ulf--G.~Mei{\3}ner, Vladimir Pascalutsa and Rocco Schiavilla for 
sharing their insights and stimulating discussions. This work has been supported by the 
U.S.~Department of Energy Contract No.~DE-AC05-84ER40150 under which the 
Southeastern Universities Research Association (SURA) operates the Thomas Jefferson 
National Accelerator Facility

\setlength{\bibsep}{0.2em}
\bibliographystyle{h-physrev3.bst}
\bibliography{/home/evgeny/refs}

\end{document}